\documentclass{iau}
\usepackage{graphicx}

\title[star clusters with top-heavy IMF] 
{Dynamical evolution of star clusters with top-heavy IMF}

\author[Haghi et al.]   
{Hosein Haghi$^1$, Ghasem Safaei$^1$, Akram H. Zonoozi$^{1, 2}$
 \and Pavel Kroupa$^2$}

\affiliation{$^1$ Department of Physics, Institute for Advanced Studies in Basic Sciences (IASBS), Zanjan 45137-66731, Iran \\ email: {\tt haghi@iasbs.ac.ir} \\[\affilskip]
$^2$Helmholtz-Institut f\"ur Strahlen-und Kernphysik (HISKP), Universit\"at Bonn, Rheienische Friedrich-Wilhelms Universit\"at Nussallee 14-16, Bonn, D-53115, Germany}

\pubyear{2019}
\volume{351}  
\jname{Star Clusters: From the Milky Way to the Early Universe}
\editors{A. Bragaglia, M. Davies, A. Sills \& E. Vesperini, eds.}
\begin{document}

\maketitle

\begin{abstract}
Several observational and theoretical studies suggest that the initial mass function (IMF) slope for massive stars in globular clusters (GCs) depends on the initial cloud density and metallicity, such that the IMF becomes increasingly top-heavy with decreasing metallicity and increasing gas density of the forming object. Using N-body simulations of GCs starting with a top-heavy IMF and undergoing early gas expulsion within a Milky Way-like potential, we show how  such a cluster would evolve. By varying the degree of top-heaviness, we calculate the dissolution time and the minimum cluster mass needed for the cluster to survive after 12 Gyr of evolution. 
\end{abstract}

\firstsection 
\section{Introduction}

Globular clusters (GCs) lose stars as a result of stellar and dynamical evolution. While the external effects caused by the gravitational field of the host galaxy might ultimately destroy GCs, the internal effects such as stellar evolution, stellar mass segregation, stellar remnants, and  binaries can remove stars from GCs in a more subtle way, ultimately altering their dynamical structure and providing escapee stars that contribute to the host galaxy. The initial condition of a star cluster determines its fate  and the stars it loses to their host galaxy. 

The initial mass function (IMF) of stars within an embedded cluster is one of the most important initial conditions that plays a significant role in star cluster evolution. Most studies of resolved stellar populations in the disk of the Milky Way showed that stars form following an IMF that has a universal form (\cite{Kroupa_2001}, \cite{Kroupa_2002}; \cite{Kroupa_2012}) which is referred to as the ``canonical" IMF. This poses a problem for star formation theories which predict a dependence on the environment where star formation takes place. The shape of the stellar IMF of a star cluster near its upper mass limit is a focal topic of investigation as it determines the high-mass stellar content and hence the dynamics of the cluster at its embedded phase. Observations indicate that the IMF may depend on the star formation environment (cloud density and metallicity) and becomes top-heavy (i.e., overabundant in high-mass stars)  under extreme starburst conditions (\cite{Dabringhausen_2009}, \cite{Dabringhausen_2012}). The data suggest that the IMF-slope is flatter in denser and metal-poorer environments. The need for  a top-heavy IMF also has been put forward to explain the observed trend of metallicity and $M/L$ ratio found among $M31$ GCs, which shows a discrepancy with SPS prediction (\cite{Zonoozi_2016} and \cite{Haghi_2017}). If this is the case,  the evolution of GCs can significantly change due to the different mass loss rates and number of black holes (BHs) formed and consequently affects our understanding of GC survival. While many aspects of the residual-gas expulsion in the embedded cluster with the universal IMF have already been studied in the literature (\cite{Baumgardt_2003}, \cite{Baumgardt_2007}), nobody has so far tried to perform a systematic study of the survival rate of the star clusters evolving under residual-gas expulsion with a varying IMF. Here, we investigate the pure effect of the top-heaviness of the IMF on the dissolution rate of GCs. We calculate a grid of models over a  wide range of stellar masses, that undergo primordial residual-gas expulsion, and have a varying the MF-slope in the high-mass range. 

\section{The models} \label{sec:Initial Conditions}

We ran simulations with different initial cluster masses in the range $M_{i}=10^4$ to $3\times 10^5 M_{\odot}$.  For the initial density profile of the clusters, we consider the Plummer profiles in virial equilibrium.  The corresponding size scale  was set by the initial $3D$ half-mass radius, $r_{hi}$, following \cite{MarksKroupa_2012} as $r_{hi}=0.1 \times \left(\frac{M_{ecl}}{M_{\odot}}\right)^{0.13}$pc. Therefore the initial $3D$ half-mass radii of all models vary from 0.33 to 0.52 pc. We use the $N$-body code  \textsc{nbody6} (\cite{Aarseth_2003} and \cite{Nitadori2012}) to evolve model star clusters from zero age to 12 Gyr, over a range of  initial half-mass radii, embedded cluster mass, and stellar IMF-slope in the  high-mass range (i.e., $M > 1M_{\odot}$).  The dissolution time, $T_{diss}$, is defined to be the time when 95\% of the initial number of the stars are lost from the cluster.  All clusters move on circular orbits with circular velocity $V_G=220 km s^{-1}$ at galactocentric radius $R_G = 8.5$ kpc through an external galaxy that follows an NFW potential.  The range of stellar masses was chosen to be from 0.08 to 100$M_{\odot}$ and the metallicity of the clusters is Z$ = 0.001$\,dex.  Gas expulsion is assumed to start at a certain time, $t_D =0.6$ Myr. The star formation efficiency is assumed to be equal to $\epsilon = 0.33$ in all models. The gas is not simulated directly, instead, its influence on the stars is modeled as a modification to the equation of motion of stars. The stellar IMF is adopted in the form of a three-segment power-law function as follow
\begin{eqnarray}
\xi (m) \propto m^{ -\alpha}: \left\{
\begin{array}{ll}
\alpha _1 = 1.35 \hspace{0.25 cm} ,\hspace{0.25 cm} 0.08 <\frac{m}{M_\odot}<0.50\\
\alpha _2 = 2.35 \hspace{0.25 cm} ,\hspace{0.25 cm} 0.50 <\frac{m}{M_\odot}<1.00\\
\alpha _3 \hspace{1.25 cm} ,\hspace{0.35 cm} 1.0 <\frac{m}{M_\odot}<100.0
\end{array}
\right.
\end{eqnarray}

In order to cover the canonical and top-heavy IMF, $\alpha _3$ is varied from 1.5 to 2.3.

\section{The dissolution time of the star clusters}

\begin{figure*}[]
\centering
\includegraphics[width=12cm]{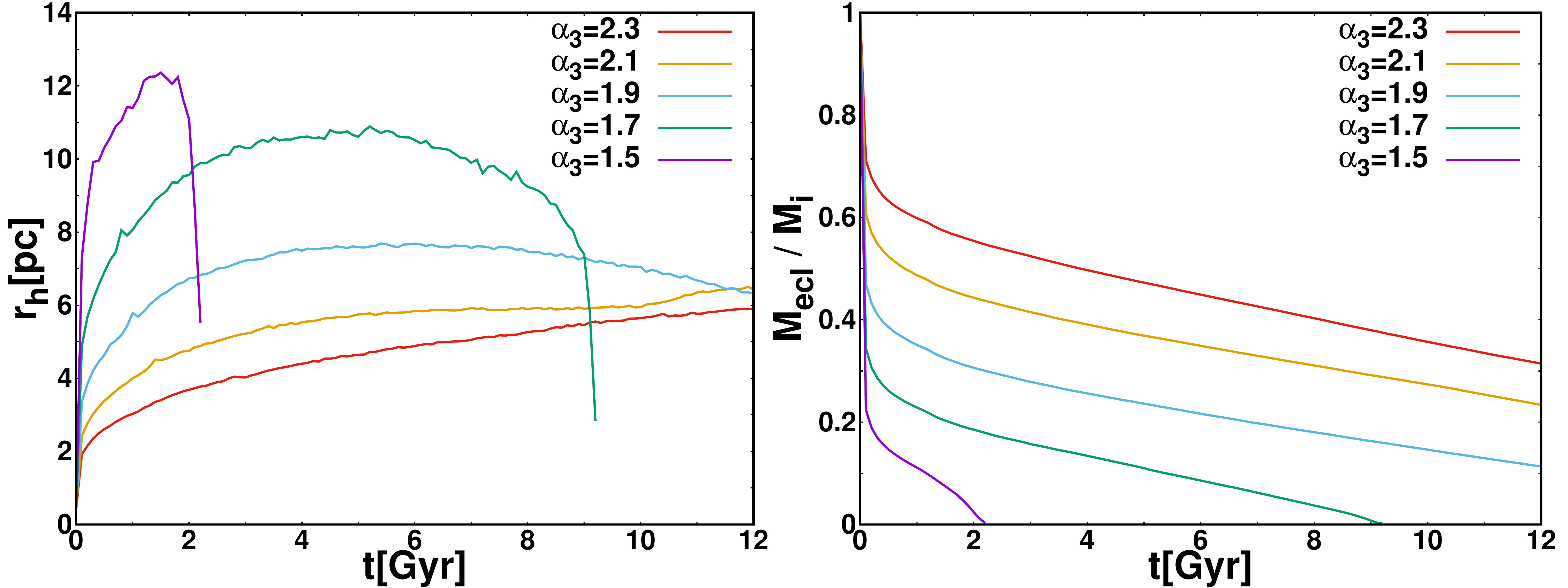}
\caption{The evolution of the 3D half-mass radius and  the total mass of a cluster with an initial mass and radius of respectively, $10^5 M_\odot$ and $r_h=0.45$ pc for different initial values for $\alpha_3$.}
\label{fig:m-rh-time-fig1}
\end{figure*}

At the beginning of the evolution, the mass-loss is dominated by gas expulsion and early mass-loss associated with stellar evolution of massive stars. All models expand violently and lose a fraction of their stars. Depending on the abundance of heavy stars in each model which is determined by $\alpha_3$,  clusters lose 30\% to 80\% of their mass within the first 100 Myr by early stellar evolution. In order to understand how the top-heaviness of the IMF (i.e., the value of $\alpha_3$) influences  the half-mass radius of bound stars  and the amount of mass lost by a star cluster during the evolution, we first consider initially identical star clusters but with a different degree of the top-heaviness. The time evolution of the total masses and half-mass radii of stars bound to the clusters are shown in Figure \ref{fig:m-rh-time-fig1}.  The initial mass and initial half-mass radius are the same but five different values are assigned to  $\alpha_3$. All models  evolve for 12 Gyr unless they are disrupted before.
We plot the initial cluster mass of $10^5 M_{\odot}$ as representative of all models. We find that increasing the degree of top-heaviness of the IMF significantly affects the evolution of the cluster. The mass loss rate is higher for the cases with a more top-heavy IMF. The cluster loses about 80\% of its mass by early rapid gas expulsion and stellar evolution in the case of the most top-heavy IMF (with $\alpha_3=1.5$), but this value decreases to 30\% for clusters with a canonical IMF (with $\alpha_3=2.3$). Consequently, the clusters with a top-heavy IMF dissolve faster than the clusters with a canonical  IMF.

\begin{figure*}[]
\centering
\includegraphics[width=67mm]{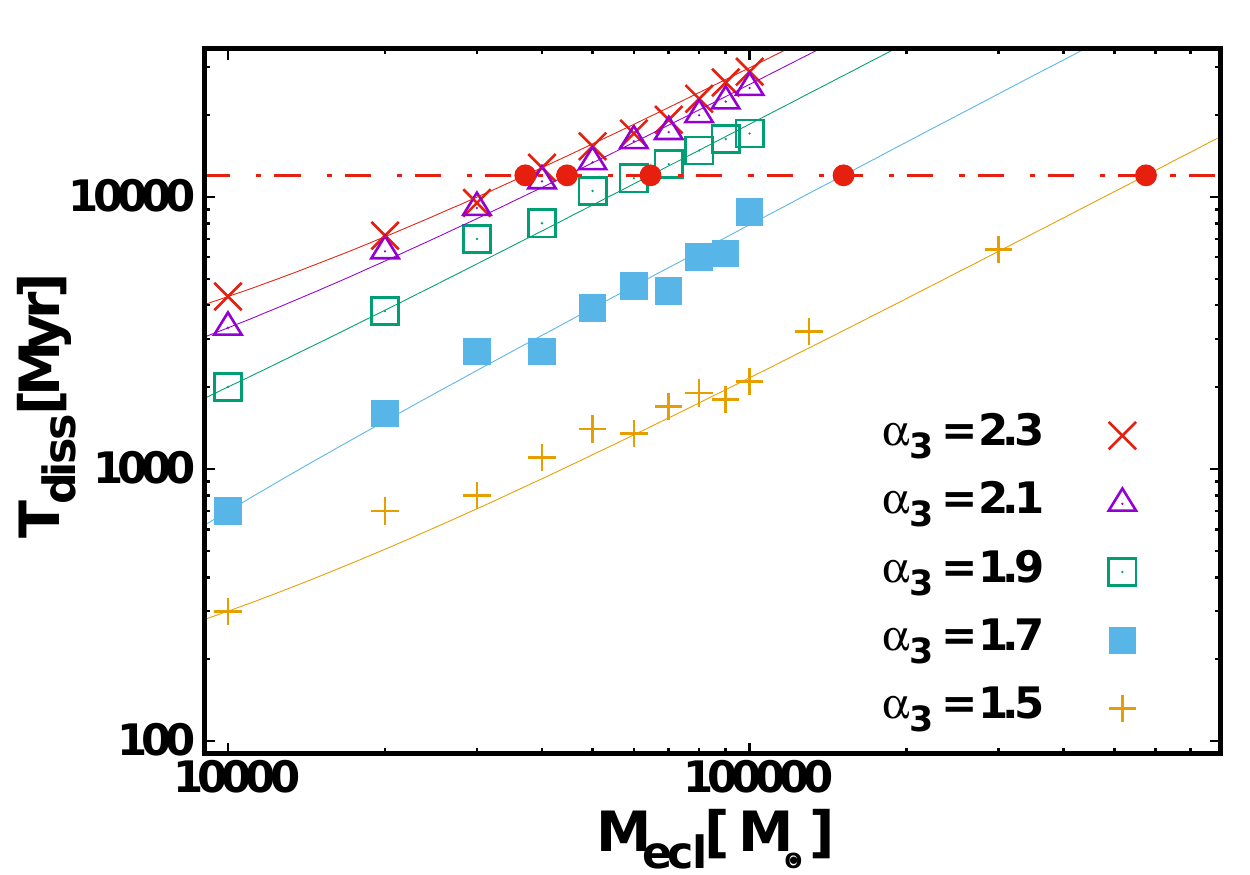}
\includegraphics[width=67mm]{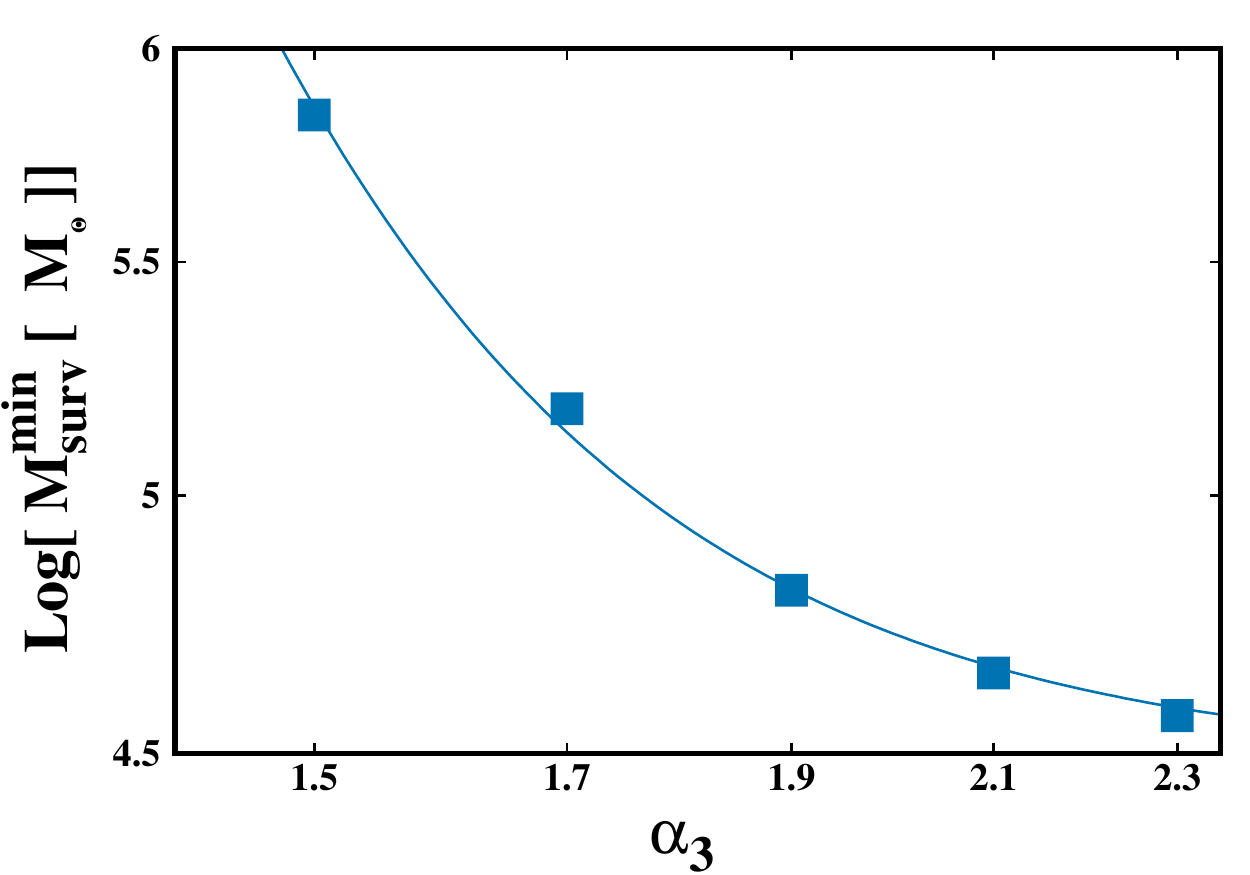}
\caption{Left: The dissolution time versus the different mass of star clusters for various $\alpha_3$.  The points show the results of N-body simulations for given initial mass and $\alpha_3$ until dissolution time. The horizontal dashed red line indicates $t=12$ Gyr.  The rising solid lines show the results of our linear fit to the results of the simulations by Equation (\ref{equ:tdiss}) for each $\alpha_3$ from 1.5 to 2.3. The filled red circles show the minimum surviving mass of the cluster for each $\alpha_3$. Right: Plotted is the minimum surviving mass of the clusters ($M^{min}_{surv}$) for various $\alpha_3$. The filled squares show the minimum mass of the clusters for every $\alpha_3$ such that clusters heavier than these points will survive after 12 Gyr. The solid line is our linear best fit. }
\label{fig:tdiss-mass-all-fig3}
\end{figure*}

Figure \ref{fig:tdiss-mass-all-fig3}  depicts the cluster lifetimes, $T_{diss}$, as a function of the initial mass in stars of the embedded cluster, $M_{ecl}$, for various IMF parameterizing with $\alpha_3$.   As can be seen, the dissolution time of clusters increases linearly with initial $M_{ecl}$. However, even though the strength of the tidal gravitational interaction due to the host galaxy is the same for both canonical and top-heavy IMF, dissolution is faster for clusters with the more top-heavy IMF (i.e. lower $\alpha_3$). This is due to the larger amount of impulsive mass-loss from stellar evolution
from clusters with the more top-heavy IMF and rapid gas expulsion which leads to a
strong expansion. We find that the dissolution time scales as
\begin{equation}
log_{10} \left[T_{diss}(\alpha_3) \right]=a(\alpha_3)~log_{10} \left[ M_{ecl} \right]+b(\alpha_3), \label{equ:tdiss}
\end{equation}
where the coefficients  $a(\alpha_3)$ and $b(\alpha_3)$ in general, depend on the degree of top-heaviness. For all models, the slopes marginally are the same within the 2$\sigma$ error bar and is nearly independent of the $\alpha_3$ parameter.  The best-fit values of $a$ and $b$ as a function of $ \alpha_3$ are  $a(\alpha_{3})=(-0.11\pm  0.13)~\alpha_{3}+(1.10\pm  0.25)$ and $b(\alpha_{3})=(1.88\pm  0.47)~\alpha_{3}-(3.98 \pm 0.91)$.  Therefore, for any system with a top-heavy IMF  it is possible to find out the  dissolution time by these functions and Eq. (\ref{equ:tdiss}).

We obtain a relation for the minimum surviving mass of the cluster ($M^{min}_{surv}$, i.e., the minimum  initial mass of the star cluster that survives within a Hubble time) as a function of $\alpha_3$. The horizontal dashed line in Fig. \ref{fig:tdiss-mass-all-fig3} shows the time $=$12 Gyr.  The intersection of the $T_{diss}-M_{ecl}$ relation for each value of $\alpha_3$  with this horizontal line (red circles) specifies the corresponding $M^{min}_{surv}$.  Determining  $M^{min}_{surv}$ might help to understand the contribution of dissolved star clusters in making and heating the Galactic thick disc.

Figure \ref{fig:tdiss-mass-all-fig3} (right panel) shows the minimum surviving mass for different values of $\alpha_3$.  By a polynomial least-square fit to the data we found that the minimum surviving mass scales  with $\alpha_3$ as $log_{10}(M^{min}_{surv}/M_\odot)=A~(\alpha_3)^{-\eta}+B$, where the coefficients $A=15.7  $ , $\eta=6$ and $B=4.5 $.   Therefore, for a given value of $\alpha_3$  one can calculate the mass of the embedded clusters that survives after a Hubble time. For example, when $\alpha_3=1.5$ the clusters heavier than $7\times 10^5 M_{\odot}$ will survive longer than  12 Gyr,  while for the canonical IMF, all clusters with a mass lower than $4\times 10^4 M_{\odot}$ will be dissolved.

\section{CONCLUSIONS}
\label{sec:conclusions}

We have performed a large grid of simulations studying the impact of varying the degree of top-heaviness with the initial half-mass radius derived from \cite{MarksKroupa_2012} and initial gas expulsion on the survival rate and final properties of star clusters. Our simulations show that the degree of top-heaviness have a strong influence on the
evolution of star clusters. All star clusters expand due to gas expulsion and for various $\alpha_3$, star clusters expand by a factor of 3 or 4.   

Open or globular clusters must, therefore, have formed more compact and with higher central densities than with what they have today. The simulations reported here should be useful for a number of follow-up projects. Our computations also allow us to study the impact of various $\alpha_3$ on the evolution of dissolution time of whole embedded cluster systems in galaxies. The mass fraction remaining bound to each individual cluster, half-mass radius, the dissolution time and the expansion rate can be calculated by interpolating between the runs in our grid.

\end{document}